\documentclass[10pt, a4paper, twocolumn]{article}
\usepackage[utf8]{inputenc}
\usepackage[toc,page]{appendix}
\usepackage{amsmath}
\usepackage{amsfonts}
\usepackage{mathabx}
\usepackage{amssymb}
\usepackage[margin=.75in]{geometry}
\usepackage{color}
\usepackage{graphicx}
\usepackage{booktabs}
\usepackage{rotating}
\usepackage[font = small]{caption}
\usepackage{subcaption}
\usepackage{graphicx}
\usepackage{abstract}
\usepackage{url}
\usepackage{cite}
\usepackage{authblk}

\newcommand{\G}{\mathrm{G}}

\author[]{Simon R. Pocock}
\author[]{Paloma A. Huidobro}
\author[]{Vincenzo Giannini}

\affil[]{Physics Department, Blackett Laboratory, Imperial College London, Prince Consort Road, London SW7 2AZ}
\date{}

\title{The effects of retardation on the topological plasmonic chain: plasmonic edge states beyond the quasistatic limit}

\begin{document}
\twocolumn[
  \maketitle
  \vspace{-7mm}
  \begin{onecolabstract}
  We study a one-dimensional plasmonic system with non-trivial topology: a chain of metallic nanoparticles with alternating spacing, which is the plasmonic analogue to the Su-Schreiffer-Heeger model. We extend previous efforts by including long range hopping with retardation and radiative damping, which leads to a non-Hermitian Hamiltonian with frequency dependence. We calculate band structures numerically and show that topological features such as quantised Zak phase persist due to chiral symmetry. This predicts parameters leading to topologically protected edge modes, which allows for positioning of disorder-robust hotspots at topological interfaces, opening up novel nanophotonics applications.
  \end{onecolabstract}
  \vspace{5mm}
]
\section*{Introduction}

Plasmonic systems take advantage of subwavelength field confinement and the resulting enhancement to create hotspots, with applications in medical diagnostics, sensing and metamaterials~\cite{Stefan,Vincenzo}. Arrays of metallic nanoparticles support surface plasmons that delocalise over the structure, and whose properties can be manipulated by tuning the dimensions of the particles and their spacing~\cite{Plasmon1,Plasmon2,Plasmon3,Abajo}. In particular $1D$ and $2D$ arrays have significant uses in band-edge lasing~\cite{Lasing1,Lasing2}, and can be made to strongly interact with emitters~\cite{Emitters1,Emitters2}. Configurations of nanoparticle dimers have been shown to exhibit interesting physical properties~\cite{Modify}; in the following we consider a nanoparticle dimer array in the context of topological photonics.

The rise of topological insulators (TIs), materials with an insulating bulk and surface states which are protected from disorder, has inspired the study of analogous photonic and plasmonic systems~\cite{TopGraph,TopGraph2,TopPlasmon1,TopPlasmon2,
TopPlasmon3,EdgeStates,TopPlasmon4,TopPlasmon5,Saba,Khanikaev,Hess,Zhang}. Topological photonics shows exciting potential for unidirectional plasmonic waveguides~\cite{QSHEL}, lasing~\cite{TopLasing}, and field enhancing hotspots with robust topological protection, which could prove useful for nanoparticle arrays on flexible substrates~\cite{PNAS1}. Plasmonic and photonic systems provide a powerful platform to examine TIs without the complication of interacting particles, and with interesting additional properties like non-Hermiticity~\cite{nonH4,Schom1,nonH1,nonH2,PT3,PT2}. The lack of Fermi level simplifies the excitation of states, and the tunability made available by the larger scale allows for the study of disorder and defects in greater depth than electronic systems~\cite{TopPhot1,TopPhot2,TopPhot3}. They also simplify the study of topology in finite systems~\cite{Gleb}.

\begin{figure}[b!]
\centering
\includegraphics[width=\linewidth]{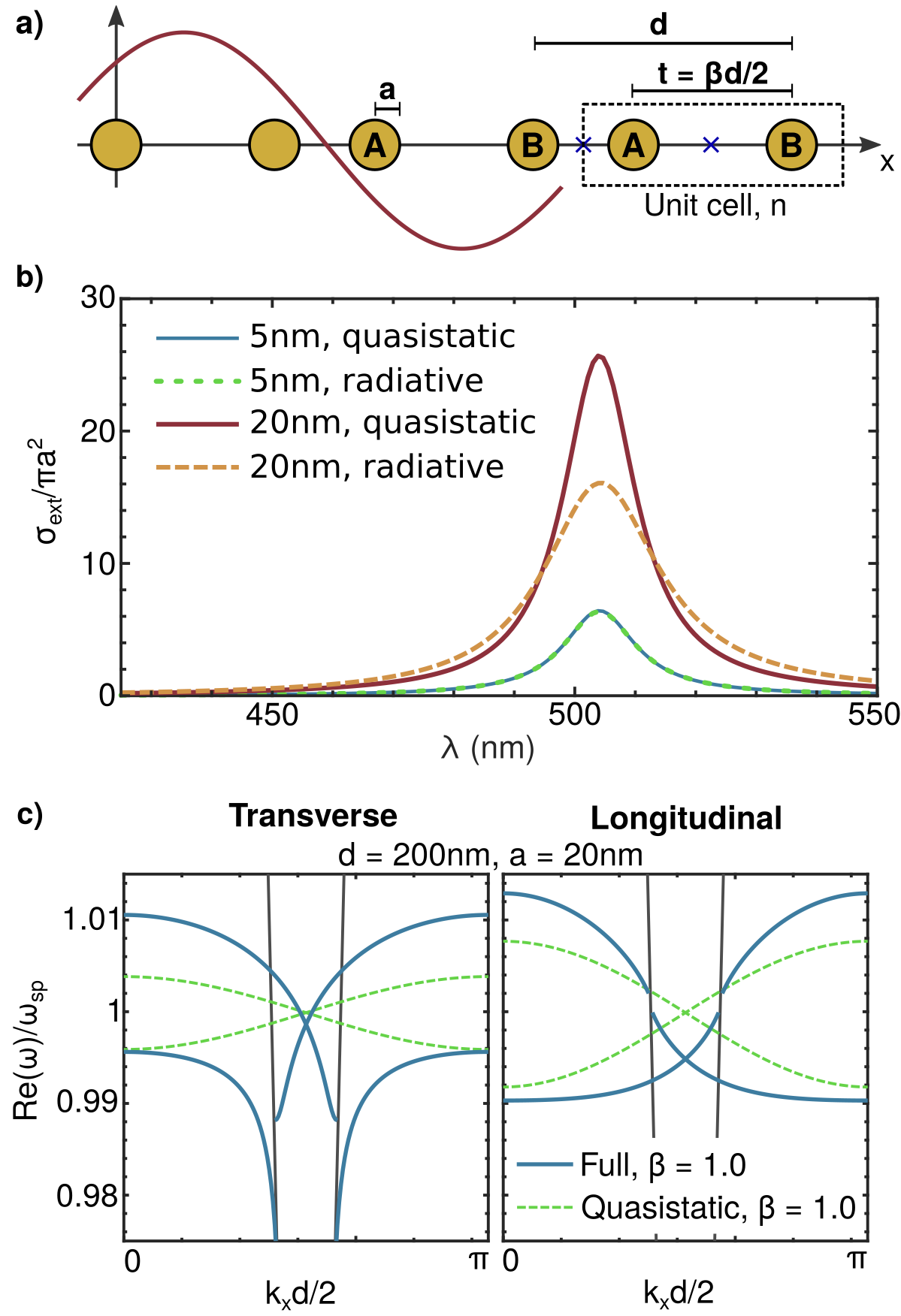}
\caption{(a) A diagram of the topological plasmonic chain, with inversion centres of the chain marked by blue crosses. (b) The extinction cross section of a single gold nanoparticle with radius $5$ and $20$~nm, comparing the effects of the QS polarizability against the radiative loss corrected polarizability. (c) Dispersion relations for an equally spaced chain ($\beta = 1$) of nanoparticles, comparing the results of the nearest neighbour QS approximation (green dashed) and treating the full Green's function (blue solid), with light lines (black). $\omega$ is normalised to the surface plasmon resonance, $\omega_{sp}$.}
\label{fig:chain}
\end{figure}

One of the simplest topologically non-trivial models is that of Su, Schrieffer and Heeger (SSH)~\cite{SSH,sCourse}, which features a chain of atoms with staggered hopping. The one-dimensional chain of metallic nanoparticles with alternating spacing (see figure~\ref{fig:chain}(a)) has been studied in analogue to this by taking the quasistatic (QS) approximation, where the dimensions of the chain are much less than the wavelength $kd \ll 1$~\cite{CTChan,Downing,PES}. In fact, when damping is neglected and only nearest neighbours considered the two systems are physically equivalent. However, the QS limit has been shown to be insufficient for describing the band structures of larger particles and spacing in equally spaced chains~\cite{Weber,Koenderink}, as it neglects the effects of retardation and radiative losses.

In the following work we present a more complete treatment of the staggered plasmonic chain which takes into account retardation and radiative effects over long range, providing a natural extension to the SSH model that complements those already in the literature~\cite{1705.06913, PT1, TopPhases}. The model breaks chiral symmetry only trivially by adding an identity term to the Bloch Hamiltonian, which is non-Hermitian with frequency dependence. We calculate band structures and compare to the QS approximation, showing that the system is indeed still topologically non-trivial because it shares eigenvectors with a chiral system. The transverse and longitudinal modes are shown to have notably different band structures, and in the transverse case it is shown that the Zak phase is not the always the same as predicted by the QS approximation. In addition, we compare two methods of calculating the Zak phase and further confirm that for inversion symmetric systems it is possible to apply Zak's original results even in the case of non-Hermiticity. We go on to study the effects of disorder on the topologically protected edge states, which we find to be extremely robust.

\section*{Topological plasmonic chain}

The plasmonic analogue of the SSH model is a chain of metallic nanoparticles with alternating spacing, as in figure~\ref{fig:chain}(a). Particles have radius $a$ and unit cells length $d$, with the spacing between the $A$ and $B$ sublattices given by $t = \beta d/2$, where $\beta$ acts as a tuning parameter. If the spacing of the particles is large enough compared to the radius of the spheres ($t,\hspace{1mm} d-t \geq 3a$) the nanospheres can be treated as dipoles with dipole moments $\mathbf{p}_n$, and the system is described by the coupled dipole equations:

\begin{equation} \label{eqn:CDE}
\frac{1}{\alpha(\omega)}\mathbf{p}_n = \sum\displaylimits_{j \neq n} \G (\mathbf{r}_{nj},\omega)\mathbf{p}_j,
\end{equation}

\noindent where $\G(\mathbf{r}_{nj},\omega)$ is the free space $3 \times 3$ Green's dyadic, which depends on the separation of the dipoles, $\mathbf{r}_{nj}=\mathbf{r}_n-\mathbf{r}_j$ and complex frequency $\omega$.

The properties of the individual nanospheres are represented by the polarizability $\alpha(\omega)$, which in the quasistatic approximation is given by,

\begin{equation}
\alpha_{qs}(\omega) = 4 \pi a^3 \epsilon_0 \frac{\epsilon(\omega) - \epsilon_B}{\epsilon(\omega) + 2\epsilon_B},
\end{equation}

\noindent where $\epsilon(\omega)$ is the dielectric function of the metal, $\epsilon_0$ the permittivity of free space and $\epsilon_B$ the background dielectric. This neglects radiative damping, which is essential for the model to be consistent with the optical theorem~\cite{Novotny}. This is addressed by the radiative correction,

\begin{equation}
\alpha(\omega) = \frac{\alpha_{qs}(\omega)}{1 - i \frac{k^3}{6\pi\epsilon_0}\alpha_{qs}(\omega)}.
\end{equation}

\noindent Throughout this work we consider gold nanospheres using an extended Drude model with $\epsilon_\infty = 9.1$, $\omega_P = 1.38\times 10^{16}$~Hz and $\gamma_D = 1.08\times 10^{14}$~Hz~\cite{Imp}, embedded in a material like glass, $\epsilon_B = 2.25$. Figure ~\ref{fig:chain}(b) shows the normalised extinction cross section of a single nanoparticle, where the solid lines are quasistatic and the dashed make use of the radiative correction. While for particles of $5$~nm radius the QS approximation agrees with the radiative correction, for particles with radius $20$~nm radiative losses strongly effect the extinction cross section by reducing and broadening the resonance over wavelength. Particles with radius as small as $5$~nm are well described classically and do not require quantum size effects to be taken into account~\cite{Jamie}.

The chain is confined to the $x$-axis, so the longitudinal ($x$) and transverse ($y$, $z$) parts of equation~\ref{eqn:CDE} decouple,

\begin{equation} \label{eqn:finite}
\frac{1}{\alpha(\omega)}p_{\nu,n} = \sum\displaylimits_{j\neq n} \G_{\nu} (r_{nj},\omega)p_{\nu,j},
\end{equation}

\noindent where $\nu = x,y,z$ labels the orientation of the dipoles, and the hopping between dipoles is scalar,

\begin{align}
\G_{x}(r_{nj},\omega) &= \frac{2e^{ikr_{nj}}}{4\pi\epsilon_0 r_{nj}^3}\left[ 1 - ikr_{nj} \right],\\
\G_{y,z}(r_{nj},\omega) &= \frac{-e^{ikr_{nj}}}{4\pi\epsilon_0 r_{nj}^3}\left[ 1 - ikr_{nj} - k^2r_{nj}^2\right],
\end{align}

\noindent with $\omega$ dependence contained within the wavenumber $k = \sqrt{\epsilon_B}\omega/c = \sqrt{k_x^2 + k_y^2 + k_z^2}$.

Unlike this work, previous studies have ignored Drude damping ($\gamma_D = 0$) and taken the QS approximation with nearest neighbour hopping to solve the system~\cite{CTChan,Downing,PES}, which neglects retardation by assuming that the dimensions of the chain are very small compared to the wavelength $kd \ll 1$. In this limit,

\begin{equation}
\G_{\nu}(r_{nj},\omega) \to m_{\nu}\frac{1}{4\pi\epsilon_0r_{nj}^3},
\end{equation}

\noindent where $m_{\nu}=2$ for the longitudinal case and $-1$ for transverse. This removes all $\omega$ dependence from $\G_\nu$ and neglects the intermediate and long range dipolar interactions. The resulting nearest neighbour, real, staggered hopping provides a close analogue to the SSH model, apart from a transformation from the eigenvalues to the frequency $\omega$.

For gold nanoparticles embedded in glass the non-radiative surface plasmon resonance $\omega_{sp} = \omega_P/\sqrt{\epsilon_\infty + 2\epsilon_B}$ corresponds to the wavelength $\lambda_{sp} = 504$~nm. Figure~\ref{fig:chain}(c) shows the dramatic difference between the QS approximation and the retarded treatment for an evenly spaced chain ($\beta = 1$) when $d=200$~nm, on the same order as $\lambda_{sp}$. Green dashed lines show bands resulting from the QS approximation, which are therefore symmetric around $\omega_{sp}$. The blue solid lines show the result of including retardation and radiative effects; as has previously been shown for the evenly spaced chain, retardation leads to polariton splitting and discontinuities at the light lines $k = \pm k_x$~\cite{Weber,Koenderink}, which are completely absent in the QS approximation. The difference is even greater for the transverse polarisation due to the extremely long range $\sim \exp(ikr)/r$ term in the full dipolar interactions. When including retardation and radiative losses hopping becomes complex and long range, giving rise to a non-Hermitian topologically non-trivial Hamiltonian.

\subsection*{Bulk Bloch Hamiltonian}

Topologically non-trivial systems exhibit a bulk-boundary correspondence, where properties of the bulk, here the Zak phase~\cite{Zak}, predict the existence or absence of edge states in the finite case~\cite{ZakGraph}. We study the bulk by way of an infinite chain, where we relabel the two particles in the unit cell $A$ and $B$ as in figure~\ref{fig:chain}(a), apply Bloch's theorem, and arrive at the equations

\begin{equation} \label{eqn:BlochH}
\mathcal{G}_{\nu}(k_x,\omega)
\begin{pmatrix}
p_{\nu}^A \\
p_{\nu}^B
\end{pmatrix}  = \frac{1}{\alpha(\omega)}
\begin{pmatrix}
p_{\nu}^A \\
p_{\nu}^B
\end{pmatrix},
\end{equation}

\noindent where $\mathcal{G}_{\nu}(k_x,\omega)$ acts as an $\omega$-dependent non-Hermitian Bloch Hamiltonian which is, in matrix form,

\begin{equation}
\begin{pmatrix}
\sum\displaylimits_{n}' \G_{\nu} (nd,\omega) e^{ik_xnd} & \sum\displaylimits_{n} \G_{\nu} (nd + t,\omega) e^{ik_xnd} \\
\sum\displaylimits_{n} \G_{\nu}(nd - t,\omega) e^{ik_xnd} & \sum\displaylimits_{n}' \G_{\nu} (nd,\omega) e^{ik_xnd}
\end{pmatrix}.
\end{equation}

\noindent The eigenvalues of $\mathcal{G}_\nu$ are $1/\alpha(\omega)$, but the band structure is given by $\omega$. Topological properties of the system are associated with the eigenvectors $\mathbf{p}_\nu$.

As is the case for any two band Hamiltonian, it is possible to write $\mathcal{G}$ in terms of the Pauli matrices $\{\sigma_i\}$,

\begin{equation} \label{eqn:pauli}
\mathcal{G}(k_x) = g_0(k_x)I + \mathbf{g}(k_x)\cdot \boldsymbol{\sigma},
\end{equation}

\noindent with $g_0$ and $\mathbf{g} = (g_x,g_y,g_z)$ given by examining $\mathcal{G}_\nu$. The QS nearest neighbour approximation has strict chiral symmetry, or sublattice symmetry, because there is no hopping from sites $A$ to $A$ or from $B$ to $B$. This can be expressed by the equation $\sigma_z\hat{H}\sigma_z = -\hat{H}$, which is true when $g_0 = 0 = g_z$. For the retarded treatment we still have $g_z = 0$, but $g_0 \neq 0$, which we will call `trivial' chiral symmetry breaking. Strict chiral symmetry leads to eigenvalues that are symmetric around $0$, but here they are symmetric around $g_0(k_x,\omega)$. This system still has inversion symmetry in the $x$-direction expressed by $\sigma_x\mathcal{G}_\nu(k_x)\sigma_x = \mathcal{G}_\nu(-k_x)$, with inversion centres marked by crosses in figure~\ref{fig:chain}(a). This guarantees that the band structure is symmetric in $k_x$.

Calculating the band structure is a matter of fixing real $k_x$ and finding corresponding complex $\omega$ numerically, to solve

\begin{equation} \label{eqn:det}
\det\left(\mathcal{G}_\nu - \frac{1}{\alpha(\omega)}I\right) = 0.
\end{equation}

\noindent This is complicated by the fact that the elements of $\mathcal{G}_\nu$ are infinite, slowly converging sums. Faster evaluation can be achieved by writing the sums in terms of polylogarithms and the Lerch transcendental function, detailed in the supporting information (SI). These analytical expressions show that when $\beta = 1$, the off-diagonals of $\mathcal{G}_\nu$ are zero at the edge of the Brilluoin zone (BZ), at $k_xd/2 = \pi/2$. Therefore the eigenvalues are degenerate here, leading to a band crossing as in figure~\ref{fig:chain}(c). This signifies a topological phase transition at $\beta = 1$.

\begin{figure}[t]
\centering
\includegraphics[width=\linewidth]{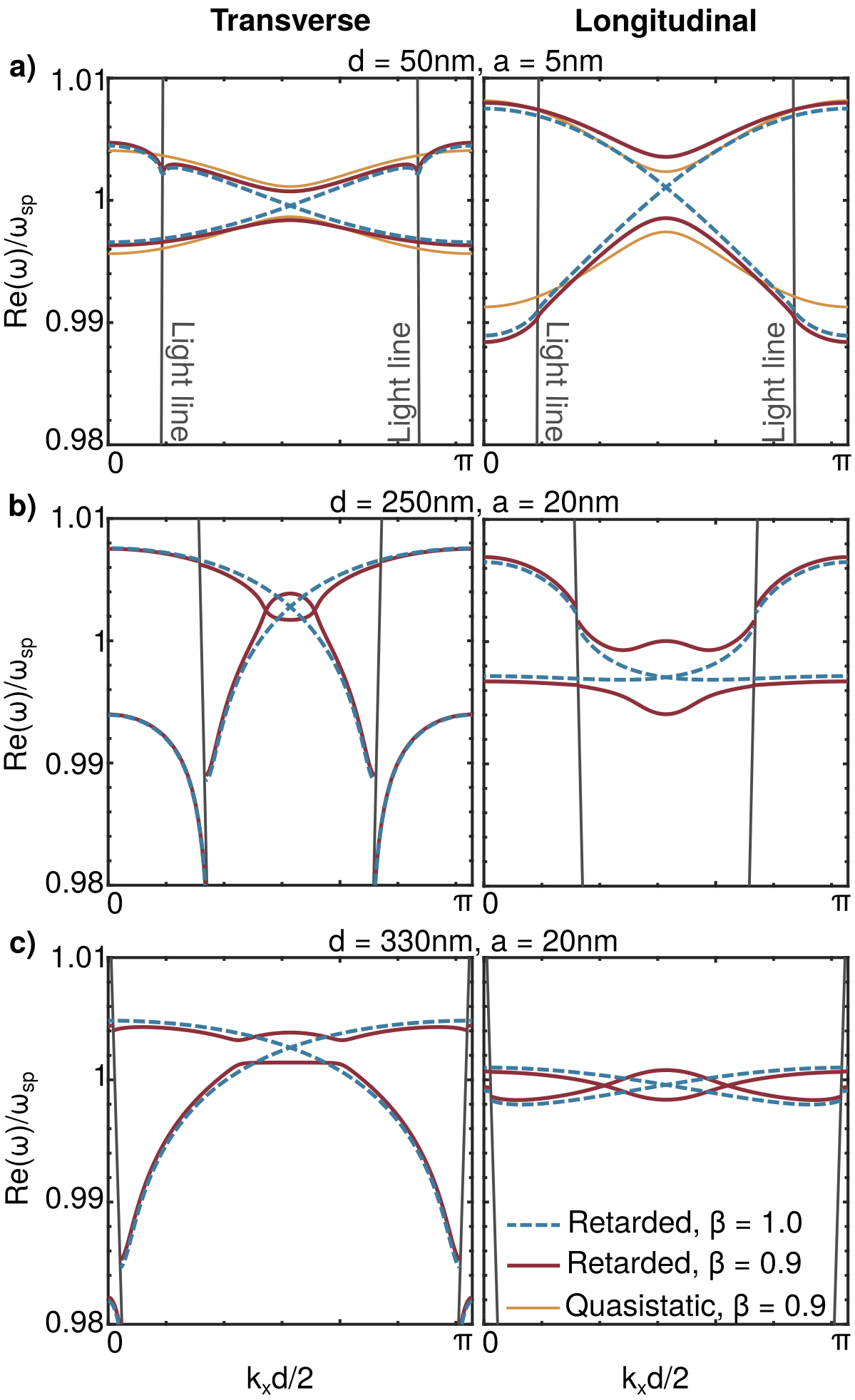}
\caption{Band structures for the topological plasmonic chain for various choices of geometric parameters $d$ and $a$, with $\beta = 0.9$ or $1.1$. (a) Comparison of QS and retarded band structures near the QS regime. (b),(c) show band structures further away from the QS regime and \emph{do not feature the yellow QS bands}, which are similar to in (a).}
\label{fig:bands}
\end{figure}

Figure~\ref{fig:bands} shows numerically calculated band structures for various choices of chain parameters $d$ and $a$, displaying only the real part of $\omega$, with some discussion of $\mathrm{Im}(\omega)$ in the SI. Blue dashed lines show the $\beta = 1$ case, and red solid lines show the $\beta = 0.9$ (identical to $\beta = 1.1$) case. For $\beta \neq 1$ a \emph{complex valued} gap opens at the edge of the BZ, which increases in magnitude with increasing $|\beta -1|$.

For small chain geometry ($d= 50$~nm, $a=5$~nm) in figure~\ref{fig:bands}(a) the band structure is well approximated by the QS model (yellow line), which makes a reasonable prediction of the band gap but fails to predict the small deviations of the band structure at the light line in the transverse case. Already some asymmetry in $\mathrm{Re}(\omega)$ exists due to the trivial breaking of chiral symmetry.

Figures~\ref{fig:bands}(b,c) demonstrate band structures with larger particles and spacing with $d\sim \lambda_{sp}$, well away from the QS limit. Once again the polariton splitting and $\mathrm{Re}(\omega)$ asymmetry are present, as well as discontinuities at the light line, as in the upper band for (b) longitudinal. The QS approximation, not shown for clarity, is poor here and completely fails to predict that a gap in $\mathrm{Re}(\omega)$ does not always open, such as in (b) transverse and (c) longitudinal. It is important to note that in these cases there still exists an gap in $\mathrm{Im}(\omega)$, such that there is no band crossing and associated topological phase transition (see SI). This means these cases can still be topological or trivial.

\subsection*{The Zak phase}

The relevent topological number is the Zak phase, which for an Hermitian system like the SSH model is given by

\begin{equation} \label{eqn:Zak}
\gamma_H = i\int\displaylimits_{-\pi/d}^{\pi/d} \mathbf{p}^\dagger \frac{\partial \mathbf{p}}{\partial k_x} \mathrm{d}k_x.
\end{equation}

\noindent However, the Hamiltonian $\mathcal{G}_\nu$ is non-Hermitian, so we must be more careful. The generalisation of the Berry phase for non-Hermitian systems~\cite{nonH1}, written in $1D$ for our system, is

\begin{equation}
\gamma = i\int\displaylimits_{-\pi/d}^{\pi/d} \mathbf{p}_L^\dagger \frac{\partial \mathbf{p}_R}{\partial k_x} \mathrm{d}k_x,
\end{equation}

\noindent where $\mathbf{p}_R$ and $\mathbf{p}_L$ are normalised biorthogonal right and left eigenvectors, solving equation~\ref{eqn:BlochH} and its Hermitian conjugate respectively. It has been shown that chiral symmetry quantises this non-Hermitian Zak phase~\cite{nonH2}. As discussed previously, our Hamiltonian breaks chiral symmetry trivially due to an additional identity term $g_0$. Since all vectors are eigenvectors of the identity, the system shares eigenvectors with a chirally symmetric counterpart (see SI) and the chiral symmetry result quantising the Zak phase applies for this system too. In fact, the inversion symmetry of the system leads to quantisation of the Hermitian Zak phase as well~\cite{TopOrig}, so that in this case both calculations have the same result,

\begin{equation} \label{eqn:phi}
\gamma = \phi(\pi/d) - \phi(0),
\end{equation}

\noindent where $\phi$ is the relative phase difference between $p^A$ and $p^B$. It follows that $\gamma$ is either $0$ or $\pi$ modulo $2\pi$. It is important to note that topological systems with chiral symmetry but no inversion symmetry do exist, so the topological nature here arises because of the (trivially broken) chiral symmetry. When the Zak phase is $\gamma = \pi$ we expect topologically protected edge modes.

\begin{figure}[t!]
\centering
\includegraphics[width=\linewidth]{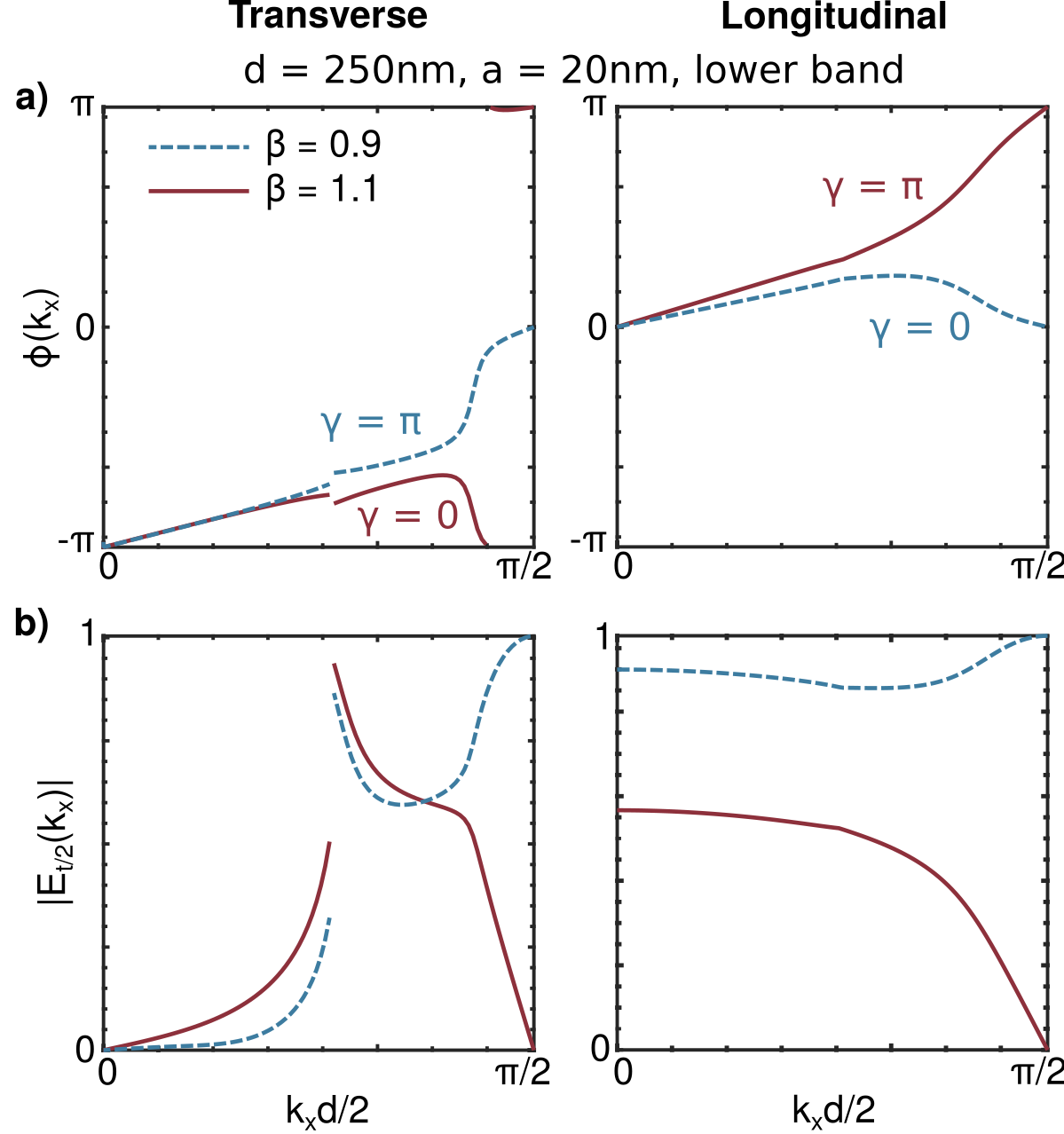}
\caption{Representations of the calculation of the Zak phase $\gamma$, considering the lower band from figure~\ref{fig:bands}(b). (a) The change of $\phi$ across the BZ modulo $2\pi$. (b) The normalised electric field at the $k_x = t/2$ inversion centre. The upper band has the same topological number in all examined cases.}
\label{fig:Zak}
\end{figure}

Figure~\ref{fig:Zak}(a) shows how $\phi(k_x)$ changes across the Brillouin zone for the lower bands of figure~\ref{fig:bands}(b). We examine $\beta$ either side of the topological phase transition at $\beta = 1$. The longitudinal case has the same property as the SSH model and QS approximation, that $\gamma = \pi$ when $\beta > 1$ and $\gamma = 0$ when $\beta < 1$~\cite{CTChan}. Surprisingly, the transverse case is in the opposite topological phase to the longitudinal case for the same choice of unit cell when it has the same $\beta$. Of the example band structures given in figure~\ref{fig:bands}, all bands have the same topological properties as the SSH model and (b) longitudinal, except (b) transverse.

Due to the existence of this unusual case and the analytically difficult elements of $\mathcal{G}_\nu$ it is unclear how to predict the Zak phase for a given set of parameters of the chain without performing the full numerics. Despite this, any interface between chains with the same $d$ and $a$, and with $\beta$ either side of the topological phase transition $\beta = 1$, will still be a topologically non-trivial interface featuring a topological edge mode.

M. Xoai \textit{et al.} showed that, in photonic systems, the Zak phase is also given by the behaviour of the electric field at the inversion centres $x = t/2, (d+t)/2$ of the chain, at the centre ($k_x = 0$) and edge ($k_x = \pi/d$) of the Brillouin zone (BZ)~\cite{photZ1}. Considering the inversion centre at $x = t/2$, if $|E_{t/2}(k_x = 0)|$ and $|E_{t/2}(k_x = \pi/d)|$ are both either zero or non-zero, we have $\gamma = 0$. If $|E_{t/2}(k_x = 0)|$ and $|E_{t/2}(k_x = \pi/d)|$ are opposite (one is zero while the other is non-zero), the Zak phase is given by $\gamma = \pi$. The normalised magnitude of the electric field at $x = t/2$ is shown across the BZ in figure~\ref{fig:Zak}(b), which agrees with the calculations of figure~\ref{fig:Zak}(a). If we take the opposite inversion centre the Zak phases are switched. This method of determining the topological nature of the system relies on Zak's results~\cite{Zak} originally for Hermitian systems, but inversion symmetry assures that any results for the Hermitian Zak phase is identical to the non-Hermitian Zak phase.

\subsection*{Finite chains and disorder}

\begin{figure}[t!]
\centering
\includegraphics[width=\linewidth]{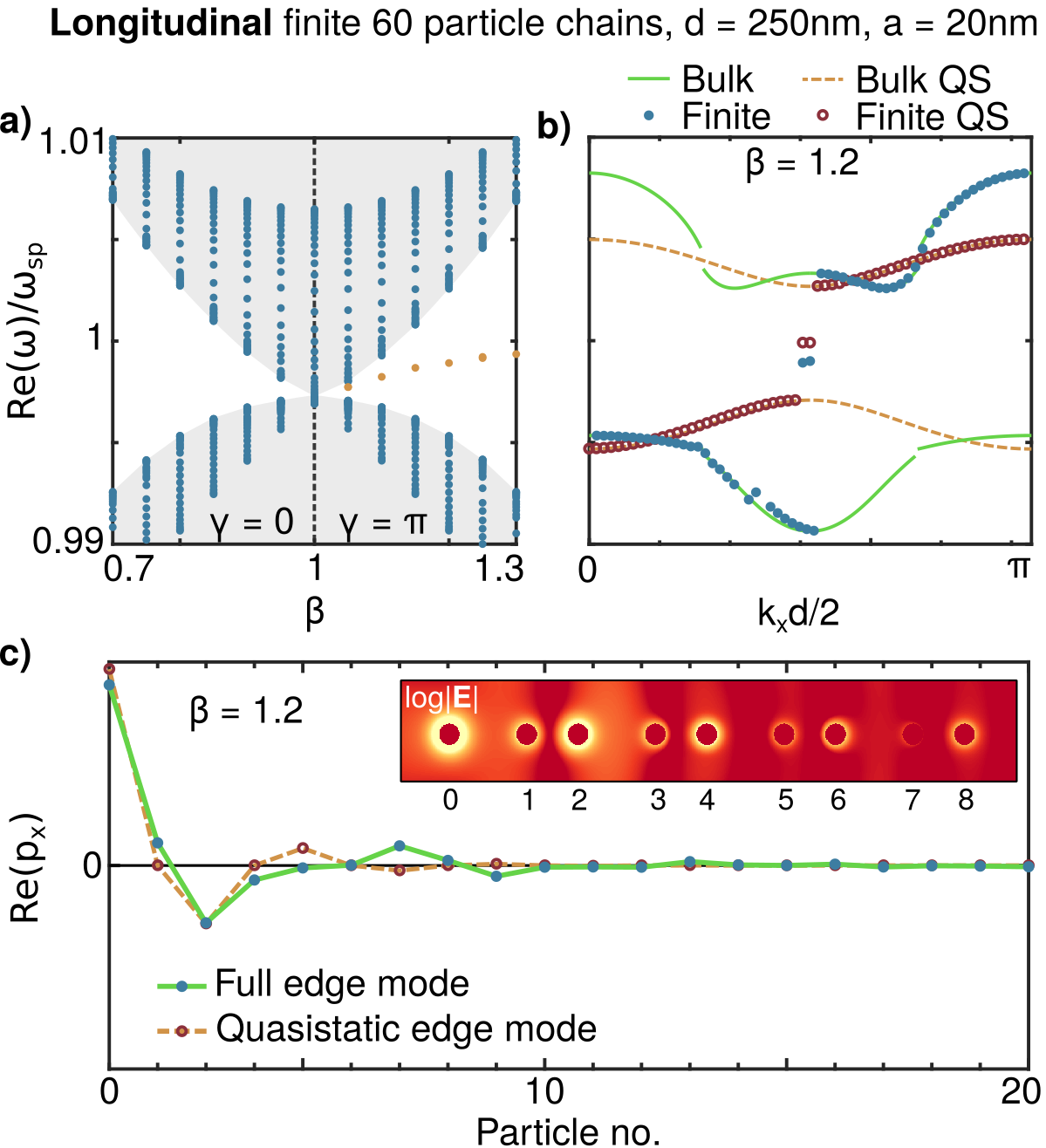}
\caption{Longitudinal. (a) Eigenmodes of a finite 60 particle chain with varying $\beta$. As predicted by the Zak phase, edge states (yellow) exist when $\beta >1$. (b) Comparison of the quasistatic and retarded finite chain band structure for a choice of $\beta = 1.2$, showing edge modes in the gap. (c) The real parts of the QS and retarded edge mode profiles of the leftmost end of the chain. Modes in the gap are symmetric and antisymmetric combinations of these modes profiles. Inset: $\log|\mathbf{E}|$ field outside of the particles for retarded left edge mode, excited by an evanescent plane wave perpendicular to the chain.}
\label{fig:Finite}
\end{figure}

We now consider the implications of the topological phases in finite chains. Figure~\ref{fig:Finite}(a) shows the eigenmodes of a finite chain with an example choice of parameters so that there is a gap in $\mathrm{Re}(\omega)$. The gap as defined by the bulk modes (blue) increases symmetrically away from $\beta = 1$, where there is a topological phase transition. As expected, edge modes (yellow) appear in the gap when the Zak phase $\gamma = \pi$.

\begin{figure}[t!]
\centering
\includegraphics[width=\linewidth]{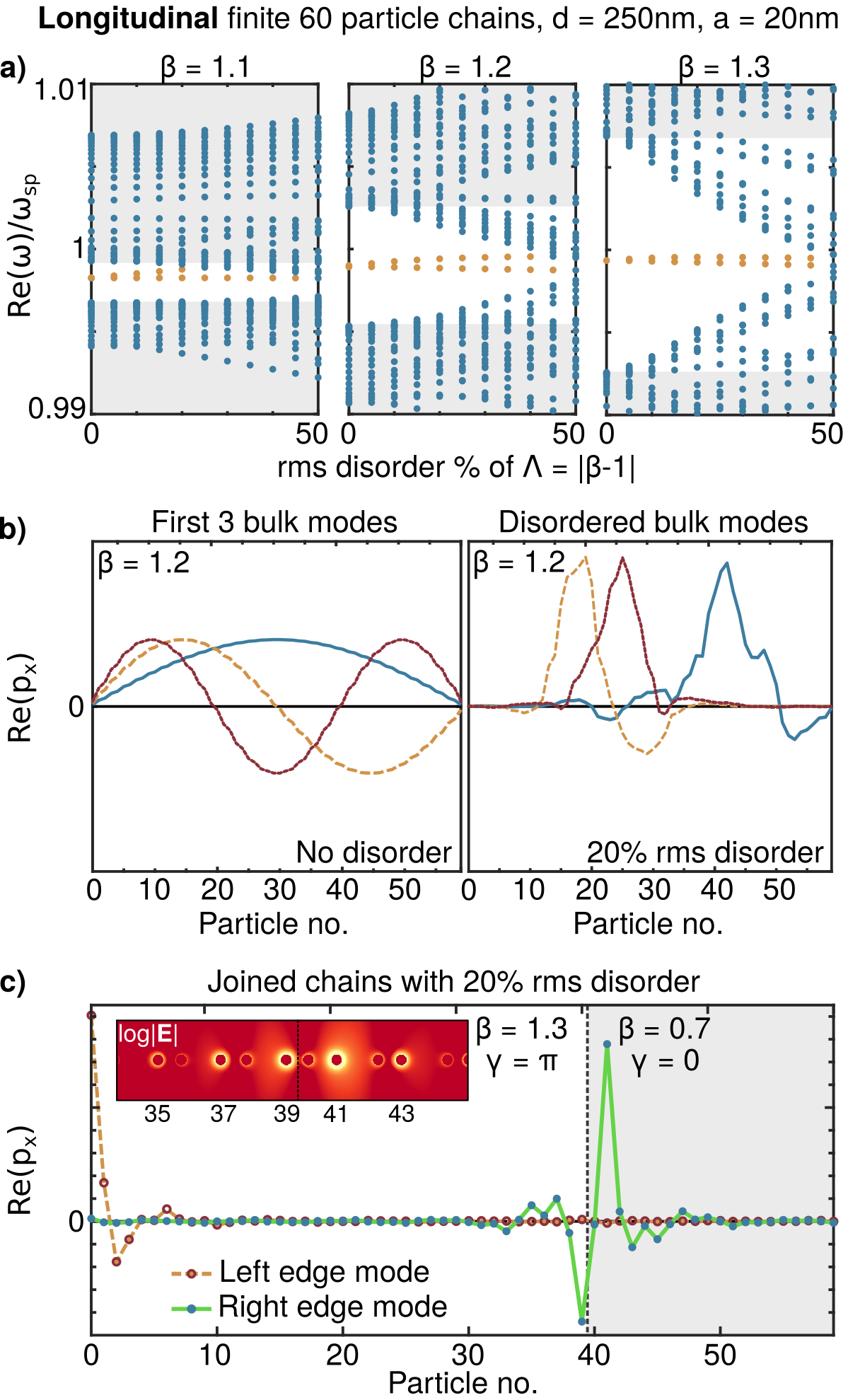}
\caption{Longitudinal. (a) Random disorder with the same seed is applied with increasing strength along the x axis for three different choices of $\beta$. The original bulk is marked in grey, bulk modes are blue and edge modes are yellow. (b) Bulk modes of the of the system without disorder and with disorder, using the same seed as (a). (c) Joined chains with opposite $|\beta-1|$ values interfaced between particles $39$ and $40$. Disorder uses a different seed to (a) and (b). Inset: $\log|\mathbf{E}|$ field outside of the particles, around the topological interface for the right edge mode excited by an evanescent plane wave perpendicular to the chain.}
\label{fig:Disorder}
\end{figure}

Bulk modes can be identified by their mode profiles, which are typically similar to normal modes of a chain, as in figure~\ref{fig:Disorder}(b). They can therefore be ordered by assigning a mode number $n$, the number of times the sign of $\mathrm{Re}(p_\nu)$ changes plus $1$, and a $k_x$ given by

\begin{equation}
k_x\frac{d}{2} = \frac{(N-2)n + 1}{N(N-1)}\pi,
\end{equation} 

\noindent where $N$ is the number of particles in the chain~\cite{Weber}, and $N=60$ in our calculations. These $(k_x,\omega)$ pairs are plotted for an example set of parameters in~\ref{fig:Finite}(b), where the finite QS approximation (red circles) and retarded system (blue dots) are compared. Bulk modes of the finite chain approximate the Bloch bulk band structure in both cases, while the two edge modes exist instead in the gap.

Figure~\ref{fig:Finite}(c) shows the dipole moments of a set of particles near the edge, with a comparison between the QS edge mode (yellow dashed) and the retarded edge mode (green solid). The QS edge mode is fully supported on only one (the A) sublattice due again to chiral symmetry, while the retarded edge mode exists on both sublattices. This is due to the long range nature of the hopping, forcing the retarded case to be further from the fully dimerised limit than the QS case. This also explains why in the QS case the edge modes have energies fixed to $\mathrm{Re}(\omega) = \omega_{sp}$ but the retarded edge modes' energies are slightly different. The QS edge mode decays exponentially into the chain whereas the retarded edge mode decreases more slowly into the chain. The real part of $p_x$ has a minimum at particle 5 before increasing and then decreasing again, because of the longer range, out of phase, dipole-dipole interactions, but the absolute value $|p_x|$ still decays monotonically on each sublattice into the chain as illustrated by the inset $\log|\mathbf{E}|$ field.

When the gap has no real part, edge mode frequencies have an imaginary part so that they still sit in the imaginary valued gap for $\gamma=\pi$ (see SI), and have similar profiles to figure~\ref{fig:Finite}(c). For the transverse case the extremely long range interactions $\sim\exp(ikr)/r$ necessitate a very long chain in order to distinguish between an edge and a bulk, which is numerically challenging.

One of the most relevant properties that arises due to topology is the protection of the edge modes from disorder in the axis of the chain. In figure~\ref{fig:Disorder} we apply disorder in the form of a random positive or negative shift to each particle's position in the chain axis, and measure the root mean square of the disorder as a percentage of $\Lambda = |\beta - 1|$. When the disorder is $50\%$ the system is within one standard deviation of the topological phase transition, where the particles are equally spaced.

In figure~\ref{fig:Disorder}(a) a random choice of disorder is scaled smoothly for different choices of $\beta$, causing the bulk modes (blue) to enter and eventually close the gap at around $50\%$. These bulk modes also become Anderson localised, as in any 1D system with disorder, with example mode profiles in figure~\ref{fig:Disorder}(b). The edge modes (yellow) separate in energy until they are lost in the bulk, but survive for high levels (sometimes $>20\%$) of disorder, especially for larger $|\beta-1|$. Figure~\ref{fig:Disorder}(c) shows the mode profiles of the two edge modes for two joined chains with $20\%$ disorder and opposite Zak phases, which illustrates the continued existence of the edge modes in disordered systems. These disorder-protected edge modes act as plasmonic hotspots, which can be positioned anywhere at the interface of two chains with opposite Zak phase.

\section*{Conclusion}

Here we have presented a detailed study of the $1D$ topological plasmonic chain beyond the quasistatic limit. We have discussed how appropriately modelling the interaction between the plasmonic nanoparticles by including the effects of retardation and radiative damping, as well as losses in the metal, leads to fundamental differences with its original electronic analogue, the SSH model. In particular, the plasmonic chain has a non-Hermitian Hamiltonian with long range hopping which breaks chiral symmetry in a `trivial' way. This implies that, because the system has the same eigenvectors as a chirally symmetric counterpart, it is still a topologically non-trivial system that supports edge modes at interfaces between topological phases.

While in the QS limit the behaviour of the SSH model is recovered, specifically band gaps opening both for transverse and longitudinal modes and edge modes confined to a single sublattice in the topological phase, we have shown that as the size of the particles and their separation increases a richer phenomenology appears. In particular, the bulk band structures deviate strongly from the QS prediction, and a band gap opening for the dimerized chain does not always appear in the real frequency axis. However, in these cases there is still a gap in the imaginary part of the frequency so that it is possible to calculate topological invariants and define topological phases. We have calculated the Zak phase and discussed a remarkable case where the topological phase of the chain is opposite to that predicted by the SSH model, suggesting that there is still greater understanding of the topological plasmonic chain to seek.

Finally, we showed that the edge states survive positional disorder in the axis of the chain to a great extent. This hints at potential uses for the $1D$ topological plasmonic chain in plasmonic systems, which could take advantage of this robustness against fabrication imperfections to design plasmonic hotspots.

\section*{Acknowledgements}

S.R.P. would like to thank Simon Lieu for many enlightening conversations about topology, and acknowledges funding from EPSRC. P.A.H. acknowledges funding from a Marie Sklodowska-Curie Fellowship.

\newpage

\section*{Supporting Information}

\subsection*{Lerch transcendent form} \label{Lerch}

One of the challenges associated with the solving of the band structure of the infinite system is how slowly the sums converge in the Bloch Hamiltonian $\mathcal{G}_\nu$, given by

\begin{equation}
\begin{pmatrix}
\sum\displaylimits_{n}' \G_{\nu} (nd,\omega) e^{ik_xnd} & \sum\displaylimits_{n} \G_{\nu} (nd + t,\omega) e^{ik_xnd} \\
\sum\displaylimits_{n} \G_{\nu}(nd - t,\omega) e^{ik_xnd} & \sum\displaylimits_{n}' \G_{\nu} (nd,\omega) e^{ik_xnd}
\end{pmatrix},
\end{equation}

\noindent where the primed sums have $n\neq 0$. Even for as many as 1000 elements of the sum calculations can be quite far off. To do this for the case of evenly spaced particles, corresponding here to the on-diagonal sums, Citrin~\cite{Citrin} and Koenderink and Polman~\cite{Koenderink} used polylogarithms defined by the sum,

\begin{equation}
\mathrm{Li}_s(z) := \sum\displaylimits_{k=1}^{\infty} \frac{z^k}{k^s}.
\end{equation}

\noindent These have the advtange of being implemented as standard in several scientific packages. It has been shown that we can write the on-diagonal terms in polylog form, 

\begin{align}
&\sideset{}{'}\sum\displaylimits_{n \in \mathbb{Z}} \mathrm{G}_{x}(n d)e^{i k_x n d} \nonumber \\
&= \frac{2}{4\pi \epsilon_0 d^3} \bigg\{ \mathrm{Li}_3(e^{i(k - k_x)d}) + \mathrm{Li}_3(e^{i(k + k_x)d}) \nonumber \\
&- i k d \left[ \mathrm{Li}_2(e^{i(k - k_x)d}) + \mathrm{Li}_2(e^{i(k + k_x)d}) \right] \bigg\},
\end{align}

\noindent and

\begin{align}
&\sideset{}{'}\sum\displaylimits_{n \in \mathbb{Z}} \mathrm{G}_{y,z}(n d)e^{i k_x n d} \nonumber \\
&= - \frac{1}{4\pi \epsilon_0 d^3} \bigg\{ \mathrm{Li}_3(e^{i(k - k_x)d}) + \mathrm{Li}_3(e^{i(k + k_x)d}) \nonumber \\ 
&- i k d \left[ \mathrm{Li}_2(e^{i(k - k_x)d}) + \mathrm{Li}_2(e^{i(k + k_x)d}) \right] \nonumber \\ 
&- k^2 d^2 \left[ \mathrm{Li}_1(e^{i(k - k_x)d}) + \mathrm{Li}_1(e^{i(k + k_x)d}) \right] \bigg\}.
\end{align}

The off diagonal terms, with $x_n = nd \pm t$, no longer fit the form of a polylogarithm. We resort to a more general special function, the Lerch transcendent,

\begin{equation}
\Phi(z,s,\nu) := \sum\displaylimits_{k=0}^{\infty} \frac{z^k}{(k+\nu)^s}.
\end{equation}

\noindent For our numerical calculations we make use of the fact that the Lerch transcendent can be expressed as an integral which converges much faster than the sum,

\begin{align}
\Phi(z,s,\nu) = \frac{1}{\Gamma(s)}\int\displaylimits_0^\infty \frac{t^{s-1} e^{-\nu t}}{1-ze^{-t}}\mathrm{d}t,
\end{align}

\noindent where $\mathrm{Re}(\nu) > 0$, and either $|z| \leq 1, z \neq 1,  \mathrm{Re}(s) > 0$ or $z = 1, \mathrm{Re}(s)>1$~\cite{Table}. This imposes the restriction that it's not possible to calculate the Lerch transcendent on the light lines for the transverse case, as can be seen from the following equations. 

The off diagonal elements in terms of the Lerch transcendent are

\begin{align} \label{LerchLong}
\sum\displaylimits_{n \in \mathbb{Z}} \mathrm{G}_{x}(nd \pm t)e^{i k_x nd }& \nonumber \\
= \frac{2}{4\pi \epsilon_0 d^3} \bigg\{ e^{ikt}&\Phi \left(e^{i(k \pm k_x)d},3,\frac{t}{d} \right) \nonumber \\ 
+ e^{-ikt}e^{i(k\mp k_x)d}&\Phi \left(e^{i(k \mp k_x)d},3,1-\frac{t}{d} \right) \nonumber \\ 
-ikd \bigg[ e^{ikt}&\Phi \left(e^{i(k \pm k_x)d},2,\frac{t}{d} \right) \nonumber \\ 
+ e^{-ikt}e^{i(k\mp k_x)d}&\Phi \left(e^{i(k \mp k_x)d},2,1-\frac{t}{d}  \right)\bigg]\bigg\},
\end{align}

\noindent and

\begin{align} \label{LerchTrans}
\sum\displaylimits_{n \in \mathbb{Z}} \mathrm{G}_{y,z}(nd \pm t)e^{i k_x nd}& \nonumber \\
= -\frac{1}{4\pi \epsilon_0 d^3} \bigg\{ e^{ikt} &\Phi \left(e^{i(k \pm k_x)d},3,\frac{t}{d} \right) \nonumber \\ + e^{-ikt}e^{i(k\mp k_x)d}&\Phi \left(e^{i(k \mp k_x)d},3,1-\frac{t}{d} \right) \nonumber \\ -ikd \bigg[ e^{ikt}&\Phi \left(e^{i(k \pm k_x)d},2,\frac{t}{d} \right) \nonumber \\ + e^{-ikt}e^{i(k\mp k_x)d}&\Phi \left(e^{i(k \mp k_x)d},2,1-\frac{t}{d}  \right)\bigg] \nonumber \\ -k^2 d^2 \bigg[ e^{ikt}&\Phi \left(e^{i(k \pm k_x)d},1,\frac{t}{d} \right) \nonumber \\ + e^{-ikt}e^{i(k\mp k_x)d}&\Phi \left(e^{i(k \mp k_x)d},1,1-\frac{t}{d}  \right)\bigg]\bigg\}.
\end{align}

\noindent Substituting $k_x = \pm \pi/d$ and $t = d/2$ we see from the above equations that 

\begin{equation}
\sum\displaylimits_{n \in \mathbb{Z}} \mathrm{G}(nd \pm d/2)e^{i n \pi /2 } = 0,
\end{equation}

\noindent which predicts the closing of the gap at $k_x = \pm \pi/d$, $t= d/2$ since $g_x = g_y = 0$ in equation~\ref{eqn:pseudospinor}.

\subsection*{Trivial chiral symmetry breaking} \label{TCSB}

As is the case for any $2\times2$ matrix, it is possible to write $\mathcal{G}$ in the form

\begin{equation} \label{eqn:pauliSI}
\mathcal{G}(k_x) = g_0(k_x)I + \mathbf{g}(k_x)\cdot \boldsymbol{\sigma},
\end{equation}

\noindent where $\boldsymbol{\sigma}$ is the vector of Pauli spin matrices, and for this system in particular

\begin{align} \label{eqn:pseudospinor}
g_0(k_x) &= \sum\displaylimits_{n \neq 0} \G (nd) e^{ik_x nd}, \nonumber \\
g_x(k_x) &= \left[\sum\displaylimits_{n} \G (nd + t) e^{ik_x nd}+\sum\displaylimits_{n} \G (nd - t) e^{ik_xnd} \right], \nonumber \\
g_y(k_x) &= i\left[\sum\displaylimits_{n} \G (nd + t) e^{ik_xnd}-\sum\displaylimits_{n} \G (nd - t) e^{ik_xnd} \right], \nonumber \\
g_z(k_x) &= 0.
\end{align}

Chiral symmetry or sublattice symmetry exists when there are no interactions between sites $A$ to $A$ or $B$ to $B$, which can be expressed in the form $\sigma_z\hat{H}\sigma_z = -\hat{H}$~\cite{sCourse}. This is the case when $g_0 = 0 = g_z$ in equation~\ref{eqn:pauliSI}. In this model we break this chiral symmetry in a `trivial' way. Interactions between $A$ to $A$ and $B$ to $B$ are introduced, but the $A$ and $B$ sublattices remain indistinguishable, because the difference between the retarded plasmonic system and a chirally symmetric system is an identity term $g_0 \neq 0$. This is important because it means that the system has the same eigenvectors as a chirally symmetry system with Bloch Hamiltonian $\mathcal{G} - g_0I$,

\begin{align}
\mathcal{G}\mathbf{p} &= \frac{1}{\alpha}\mathbf{p}, \nonumber \\
\left(\mathcal{G} - g_0I\right)\mathbf{p} &= \left(\frac{1}{\alpha} - g_0\right)\mathbf{p}.
\end{align}

\noindent Since the eigenvectors are the same we can apply Zak phase results about chirally symmetric systems to our trivially non-chiral system. We also have a pseudo-chiral equation for $\mathcal{G}$ from the original chiral equation, $\sigma_z (\mathcal{G} - g_0I) \sigma_z = -(\mathcal{G} - g_0I)$.

This explains why the eigenvalues are not symmetric around $1/\alpha = 0$, and goes one step further:

\begin{align}
\left(\mathcal{G} - g_0I\right)\sigma_z\mathbf{p} &= -\sigma_z\left(\mathcal{G} - g_0I\right)\mathbf{p} \nonumber \\
&= -\left(\frac{1}{\alpha}-g_0\right)\sigma_z\mathbf{p},
\end{align}

\noindent so that much like the chirally symmetric case every eigenmode $\mathbf{p}$ has a counterpart $\sigma_z\mathbf{p}$, with eigenvalues related by

\begin{align}
\mathcal{G}\mathbf{p} &= \frac{1}{\alpha}\mathbf{p}, \nonumber \\
\mathcal{G}\sigma_z\mathbf{p} &= \left(2g_0 - \frac{1}{\alpha}\right)\sigma_z\mathbf{p}.
\end{align}

\noindent These equations relate the upper and lower bands of $\mathcal{G}$, although an additional non-trivial mapping is required from these eigenvalues to $\omega$. Notably $\mathcal{G}$ also has $\omega$ dependence, which is why the eigenvalues aren't perfectly symmetric around $g_0$.

This can also be used to explain the fact that the edge modes of our system do not appear to be supported on only one of the sublattices, which is the case for the SSH model. Here the sublattice projection operators are given by $\hat{P}_A = (I + \sigma_z)/2$ and $\hat{P}_A = (I - \sigma_z)/2$. In the case of the fully dimerised SSH model edge modes $|\psi_n\rangle$ are zero energy $E_n = 0$, so that

\begin{equation}
\hat{H}\hat{P}_{A/B}|\psi_n\rangle = \hat{H}\left(|\psi_n\rangle \pm \sigma_z|\psi_n\rangle\right) = 0.
\end{equation}

An equivalent equation for our model would be

\begin{align}
\mathcal{G}\hat{P}_{A/B}\mathbf{p} &= \mathcal{G}\left(\mathbf{p} \pm \sigma_z\mathbf{p}\right) \nonumber \\
&= \frac{1}{\alpha}\mathbf{p} \pm \left(2g_0 - \frac{1}{\alpha}\right)\sigma_z\mathbf{p}.
\end{align}

\noindent Clearly $\hat{P}_{A/B}\mathbf{p}$ is only an eigenvector of $\mathcal{G}$ when $\mathbf{p}$ and $\sigma_z\mathbf{p}$ have the same eigenvalues- which we would expect in a `fully dimerised limit', except that such a thing doesn't exist due to the long range dipole-dipole interactions of the model. This suggests that the nature of the hopping makes it difficult to see a `purely' chiral behaviour, and we can expect stronger confinement to one sublattice when $|\beta-1|$ is larger.

It is possible to break chiral symmetry in a non-trivial way by adding a term in $\sigma_z$ to the Bloch Hamilotnian, for example by changing the relative sizes of $A$ and $B$ particles, or making $A$ and $B$ different metals.

\subsection*{Complex valued frequencies} \label{Freq}

\begin{figure}[htb]
\centering
\includegraphics[width=1.0\linewidth]{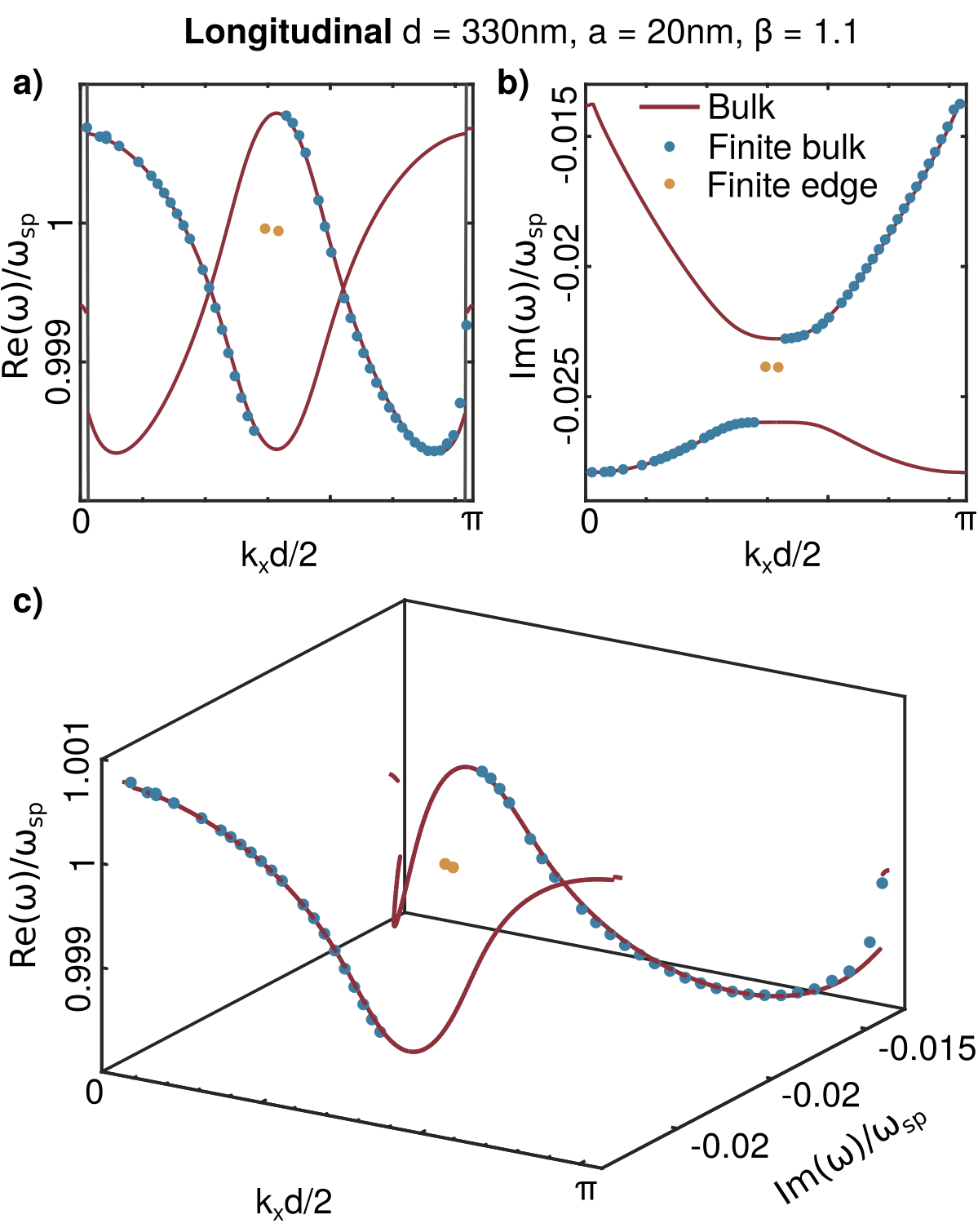}
\caption{Bulk and finite 60 particle chain band structure for $d = 330$~nm and $a = 20$~nm, $\beta = 1.1$. This is in the $\gamma = \pi$ topological phase, leading to edge modes. (a) The band structure projected onto the $\mathrm{Re}(\omega)$ axis, with finite bulk modes (blue) following the Bloch (red) bands, and edge modes with difference frequency. (b) The imaginary part of the band structure, which shows that there still exists a gap in which the edge modes sit (c) A $3D$ visualisation of the band structure. Discontinuities in the bulk band structure exist at the light line.}
\label{fig:BandsReIm}
\end{figure}

Figure~\ref{fig:BandsReIm} shows the Bloch band structure (red) in the case where there is a crossing in the real frequency axis, and the gap still has an imaginary part. In this case it is still possible to calculate a Zak phase and, for a finite chain, find bulk modes (blue) which follow the bands and edge modes (yellow) which sit in the gap. It is important to note that all of the retarded and radiative band structures presented in this work have small imaginary components to the frequency. The presence of such imaginary part implies that it may be necessary to use an evanescent wave to excite the modes in practice, which could be acheived with, for example, the usual Otto or Kretschmann configurations~\cite{Otto, Kretschmann}.

\newpage
\bibliography{TopLitReview}
\bibliographystyle{unsrt}

\end{document}